\documentclass{cupconf}
\usepackage{epsfig}

  \checkfont{eurm10}
  \iffontfound
    \IfFileExists{upmath.sty}
      {\typeout{^^JFound AMS Euler Roman fonts on the system,
                   using the 'upmath' package.^^J}%
       \usepackage{upmath}}
      {\typeout{^^JFound AMS Euler Roman fonts on the system, but you
                   dont seem to have the}%
       \typeout{'upmath' package installed. cupconf.cls can take advantage
                 of these fonts,^^Jif you use 'upmath' package.^^J}%
      }
  \else
  \fi

  \checkfont{msam10}
  \iffontfound
    \IfFileExists{amssymb.sty}
      {\typeout{^^JFound AMS Symbol fonts on the system, using the
                'amssymb' package.^^J}%
       \usepackage{amssymb}%

      }{}
  \fi

  \IfFileExists{amsbsy.sty}
    {\typeout{^^JFound the 'amsbsy' package on the system, using it.^^J}%
     \usepackage{amsbsy}}
    {}

\def\kms{\ifmmode{\,\hbox{km}\,\hbox{s}^{-1}}\else{$\,$km$\,$s$^{-1}$}\fi}
\def\etal{et~al.}
\def\farcm{\hbox{$.\mkern-4mu^\prime$}}
\def\arcsec{\hbox{$^{\prime\prime}$}}

\title[Cluster-Cluster Lensing]{Cluster-Cluster Lensing in Abell 2152$^1$}

\author[J. P. Blakeslee]{John P. Blakelsee}
\affiliation{Dept.\ of Physics \& Astronomy,
Johns Hopkins University, 
Baltimore, MD 21218}
\pubyear{2001}
\volume{538}
\pagerange{1--4}

\begin{document}
\maketitle

\begin{abstract}
We discuss lensing properties of the nearby cluster Abell 2152.  Recent
work shows that Abell~2152 is actually the projection of two different
clusters, one of which is a Hercules supercluster member at $z=0.043$,
while the other is more than 3~times as distant.  The cluster centers
have a projected separation of only 2\farcm4, and our data indicate that
the foreground cluster lenses members of the background one.  We have an
ongoing program to measure the magnification of the fundamental plane in
the background cluster.  The magnification of this standard rod will
provide an estimate of the foreground cluster mass free from the
uncertainty of the mass-sheet degeneracy which affects mass estimates
based on weak shear.\vspace{-5pt}
\end{abstract}

\footnotetext{\noindent\hbox{\scriptsize
{\it $^1$The Dark Universe: Matter, Energy, and Gravity},
STScI Symposium, 2--5\,April~2001, M.\,Livio, ed.}}

\firstsection 
\vspace{-5pt}

\section{Introduction}

The Hercules supercluster is a close grouping of three rich Abell clusters
at a redshift $z\approx0.04$.  While the richness class~2 cluster
Abell$\,$2151 (the classical ``Hercules cluster'') dominates from the
standpoint of the number of galaxies, the richness~1 cluster Abell$\,$2152
is nearest to the center of the grouping.  The third super\-cluster member
is Abell$\,$2147.  All three clusters are projected within a radius of only
$\sim\,$1$^\circ$ ($\,\sim\,$2$\,h^{-1}\,$Mpc) and
have velocity dispersions of 700 to 800 \kms\ (Barmby \& Huchra 1998).

We have found that the cluster catalogued as Abell 2152 is actually the
chance alignment of two galaxy clusters: A2152 proper at $z{\,=\,}0.043$,
and a more massive background cluster at $z{\,=\,}0.134$ which we
designate A2152-B.  The centers of these two clusters are separated by
just 2\farcm4, corresponding to 84 $h^{-1}\,$kpc at $z{=}0.043$.
Our analysis of ground-based imaging and spectroscopic data
presented below indicates that the foreground cluster is strongly lensing
at least one, and likely two, members of the background cluster.

\begin{figure}%
\begin{center}\epsfxsize=13.2cm \epsffile{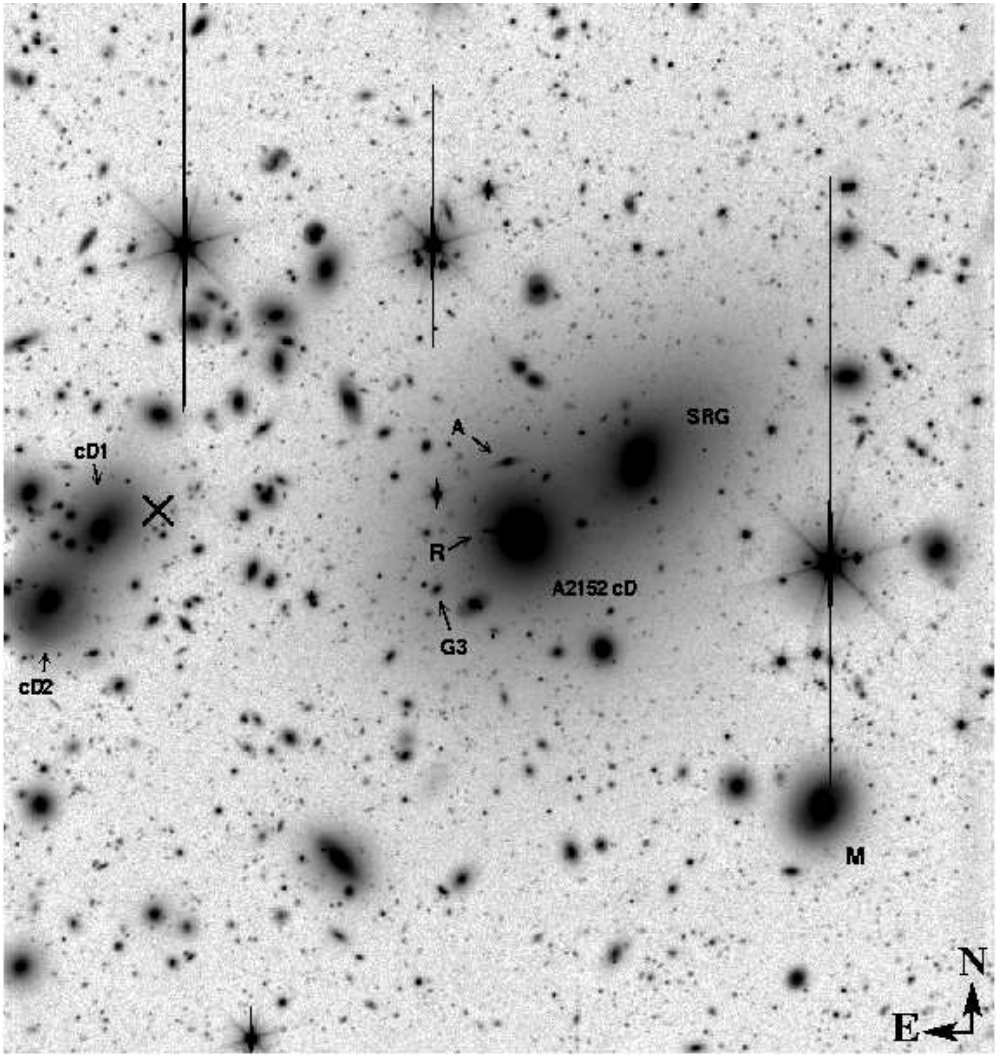}\end{center}
\caption{A 5\farcm6$\times$6\farcm0 Keck/LRIS {\it R}-band image
of A2152/A2152-B.
The bright galaxy at center is the {\it z}${\,=\,}$0.043 {\bf
A2152 cD}; the second-ranked A2152 galaxy ({\bf SRG}) is 47\arcsec\
to the northwest. The two galaxies labeled {\bf cD1} and {\bf cD2}
are in the background A2152-B cluster at {\it z}${\,=\,}$0.134.
The  $z{\,=\,}0.142$ lensed arclet ({\bf A}) appears in 
the halo of the foreground A2152 cD. A radially oriented object
({\bf R}) with an unconfirmed redshift of 0.13
may also be strongly lensed. Another confirmed
early-type member ({\bf G3}) of the A2152-B cluster happened to fall
in the slit during long-slit observations of the A2152 cD.  A large
spiral galaxy in the field with a measured redshift (labeled {\bf M})
is a member of A2152.  The position
of the X-ray center in this field is shown as a large ``{\bf X}''
near the background cD~pair.  Our $(B{-}R)$ color data reveal
that {\it most} of the smaller galaxies projected between the A2152 and
A2152-B cD's are A2152-B cluster members.  }
\label{fig:chart}
\end{figure}
\vspace{-8pt}

\section{The A2152/A2152-B Lensing System}

The magnitude $m_R=18.6$ object labeled `A' in Figure~1 is
projected 25\arcsec\ north of the A2152 cD galaxy but 
lies in the background at redshift $z{\,=\,}0.142$ (Blakeslee \etal\ 2001). 
The data were obtained at Keck Observatory using the LRIS (Oke \etal\ 1995).
The curved shape, redshift, and orientation of the object
all suggest gravitational lensing.  Such instances of low-$z$
cluster lensing in principle allow for higher resolution studies 
of the core mass profile
(e.g., Campusano \etal\ 1998; Blakeslee \& Metzger 1999).
However, our detailed modeling of this $z{=}0.043$ lensing system
required the dominant potential well to be centered on the A2152 cD,
rather than on the peak of the X-ray emission, which is positioned 2\farcm1
to the east (Jones \& Forman 1999).  Because of this, we suggested that
there was a massive background cluster surrounding a pair of bright
$z{\,=\,}0.13$ early-type galaxies, one of which is located just
20\arcsec\ from the reported X-ray center.

Recent ground-based $B$-band data confirm the existence of a rich
background cluster, which we call A2152-B. 
The ``new'' cluster shows a well-defined early-type galaxy locus
in the color-magnitude diagram of Figure~2.  The colors are
consistent with the $z=0.134$ redshift of A2152-B cD galaxy pair.
This supports our strong lensing analysis of object~`A'{\rm:}~ the A2152
potential is indeed centered on its cD, but the X-ray center is offset
because of background cluster emission.  Lensed galaxy `A' itself may well be
a high-velocity member of A2152-B.  
Another possibly strongly lensed {\it radial} feature (`R' in Figure~1)
just 11\arcsec\ from the A2152~cD may also be in A2152-B,
although its $z=0.13$ redshift is based on a
single line, and therefore unconfirmed.

The A2152/A2152-B system appears to be the first known
instance of cluster-cluster lensing.
In addition, our images reveal a faint blue arc near one of the pair
of {\it background} cD galaxies, the one labeled ``cD$\,$1'' in Figure~1.
The arc candidate is indicated in Figure~3.
If this is a yet more distant source being lensed by
the background cluster, then this would make A2152-B a {\it lensed lens}.
Additional arc candidates can be found near the background cD pair
but cannot be verified with our current ground-based data.

\begin{figure}%
\epsfxsize=9.8cm 
\begin{center}
\epsffile{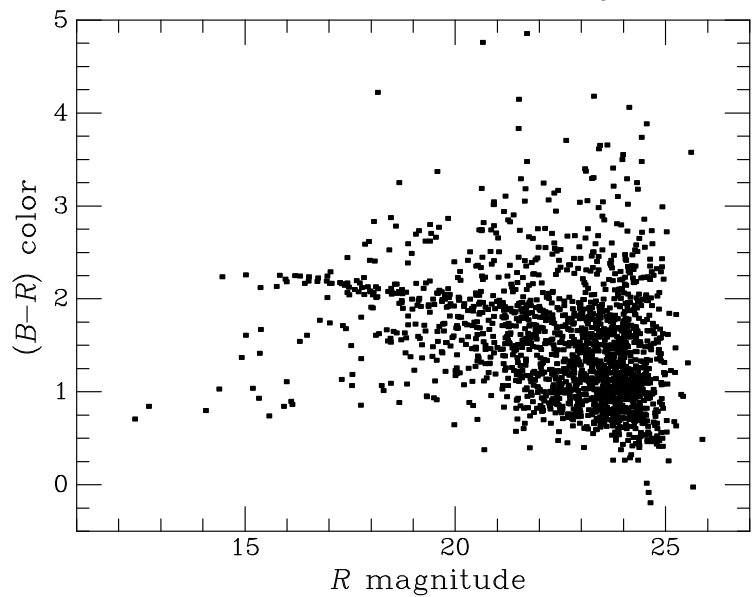}\end{center}
\caption{The $(B{-}R)$ color--magnitude diagram for the A2152 field.
The early-type galaxy population in the newly identified
$z=0.13$ cluster A2152-B stands out as a roughly horizontal
locus at $B{-}R \approx 2.1$.  The foreground $z=0.043$ cluster
locus is harder to identify because of the lower
surface density, but is shifted blueward by about 0.4~mag.}
\end{figure}

\begin{figure}%
\epsfxsize=9.3cm 
\begin{center}
\epsffile{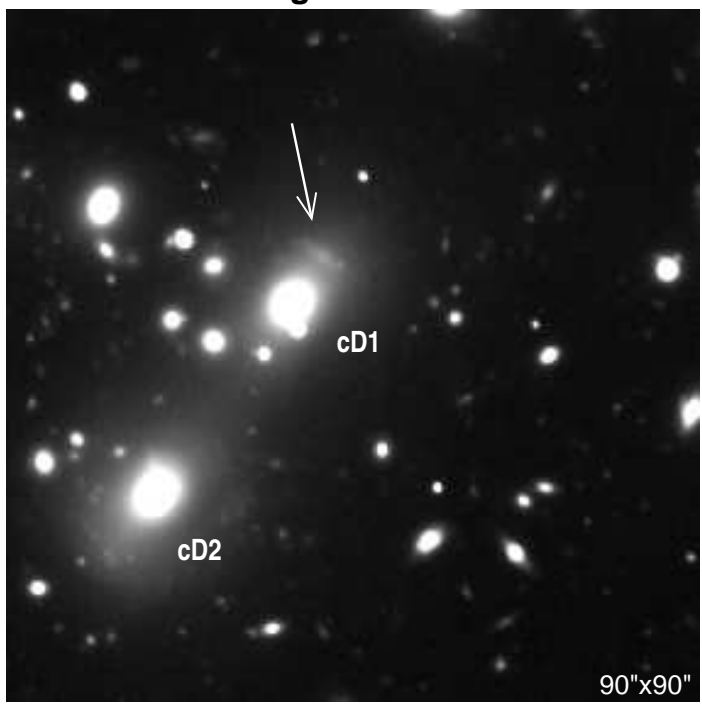}\end{center}
\caption{A blow-up of the $B$-band Keck/LRIS image
near the cD galaxy pair in the $z{\,=\,}0.134$ cluster \hbox{A2152-B}.
The possible arc in the halo of {\sc cD1}
is indicated by the arrow.  If this can be confirmed as a lensed feature,
then this would make A2152-B a lensed cluster lens.}
\end{figure}

\clearpage

\section{A Unique Opportunity}\label{sec:concl}

Because gravitational lensing measures only the total projected mass along
a line of sight, cluster masses will be overestimated {\it
whenever} there are superposed galaxy groups, sheets, or filaments.  This
is a particularly severe problem for optically selected clusters, such as
Abell clusters.  Projection of multiple mass components now appears to
explain some of the discrepancies between lensing and
X-ray/dynamical masses in several inter\-mediate redshift
clusters, including CL0024+1654 (Soucail \etal\ 2000;
Czoske \etal\ 2001) and RX J0848+4456 (Holden \etal\ 2001).
Abell$\,$2152 offers a unique opportunity to study
and disentangle a complex, overlapping cluster mass system 
at close~range.

One benefit of cluster-cluster lensing is
the large population of background galaxies at a known redshift.  
Knowing the redshift of the source population removes a
significant contribution to the uncertainty
in weak lensing reconstructions of the mass profile.  
Of course, the addition of
weak lensing information for yet more distant sources by both
clusters then helps in constraining the background cluster mass profile
as well.

Even more exciting is the opportunity to
measure the fundamental plane  in a lensed cluster.
\hbox{Because} of the lensing, the early-type galaxy population
in A2152-B should lie ``above'' the unlensed fundamental plane, i.e., 
have effective radii and luminosities
too large for their velocity \hbox{dispersions.}
The magnitude and spatial gradient of the offset will provide a
direct measure of the {\it absolute} mass profile of A2152.

We have recently obtained good quality Keck LRIS multi-slit 
spectroscopy for about 100 galaxies in the central A2152/A2152-B field.
When completed, this will be the first measurement of
gravitational magnification of a known standard candle or standard rod.
Mass estimates from magnification are unaffected by the notorious
``mass-sheet degeneracy'' which afflicts all estimates based solely on
image shear.  The combination of magnification and shear measurements at
the same projected radii should greatly enhance our understanding of the
mass structure in this complex nearby cluster lensing system.

\begin{acknowledgments}
I thank my collaborators Mark Metzger, Harald Kuntschner, and 
Pat C\^ot\'e.  I also thank Ian Smail and John Lucey for
valuable discussions and help.
\end{acknowledgments}

\end{document}